\documentclass[a4paper,aps,twocolumn,nofootinbib,nobibnotes,twoside]{revtex4-2}

\usepackage{mathrsfs}
\usepackage[scr=rsfs]{mathalfa}
 
\usepackage{array}
\usepackage{geometry}
\usepackage{amsmath}
\usepackage{amsthm}
\usepackage{bm}
\usepackage{bbm} 
\usepackage{amssymb}
\usepackage{xspace}
\usepackage{amscd}
\usepackage{dsfont}
\usepackage{t1enc}
\usepackage{scalefnt}
\usepackage[greek, english]{babel}
\usepackage[latin2]{inputenc}
\usepackage{teubner}
\usepackage{graphicx}
\usepackage{multirow}
\usepackage{dcolumn}
\usepackage{braket}
\usepackage{xifthen}
\usepackage{textcomp}
\usepackage{tikz}
\usepackage{tipa}
\usepackage{pifont}
\usepackage{cancel}
\usepackage{stackrel}
\usepackage{xcolor}
\usepackage{soul}
\usepackage{stmaryrd}
\usepackage{multibib}

\usetikzlibrary{matrix,arrows,decorations.markings,decorations.pathreplacing,petri,topaths}
\usepackage[stable,symbol]{footmisc}
\definecolor{col}{rgb}{0.25, 0.41, 0.88}
\usepackage[pdftex, pdftitle={}, pdfauthor={}, colorlinks=true, urlcolor=col, linkcolor=col, bookmarks=false, citecolor=col, raiselinks=false,hyperfootnotes=false]{hyperref}

\usepackage{actaneasy}
\usepackage[T1]{fontenc}

\begin{document}

\setcounter{pagep}{1}\setcounter{pagewst}{0}\setcounter{zkon}{0}\setcounter{page}{0}
\dess{iwobb1}
\ch{Vlasov--Klimontovich Equation in Action}
\aum{\L{}.A. Turski}
\arm{Vlasov--Klimontovich Equation in Action}
\spt{\L{}.A. Turski}{Vlasov--Klimontovich Equation in Action}
\aut{\L{}.A. Turski$^*$}{Center for Theoretical Physics, Polish Academy of Science \\ al. Lotnik\'ow 32/46, 02-668 Warsaw, Poland}
\pacc{This paper outlines a variety of possible applications of the Vlasov equation and its generalization, i.e., the Klimontovich equation, in various areas of many-body physics. In particular, these equations are shown to be used in relativistic plasma physics, the theory of semi-classical Bloch electrons, and the metriplectic description of dissipative processes.}{kinetic theory of relativistic plasma, semi-classical theory of Bloch electrons, metriplectic description of dissipative processes\vs*{0pt}}{laturski@cft.edu.pl}

\section{Introduction\vs*{4pt}}

In 1938, Anatoly Vlasov wrote a seminal paper~[1] in which he argued that in the description of a many-body system with long-range interparticle interactions, the conventional kinetic Boltzmann equation is inadequate and should be replaced by the continuity equation for the one-particle distribution function $f(\we{r},\we{p},t)$ in the one-particle phase space, where $\we{r}$ denotes the particle position and $\we{p}$~its momentum. The adequate equation in question is
\beq 
\aquad \frac{\partial f(\we{r},\we{p},t)}{\partial t}+\frac{\we{p}}{m}\cdot\nabla f(\we{r},\we{p},t)+\we{F}\cdot\frac{\partial f(\we{r},\we{p},t)}{\partial \we{p}}=0. \eeq{}\tyl \beq 
\eeq{1} 
Here, $\we{F}$ is the total force acting on the particle, resulting from interactions with all particles in the system. Therefore, the force $\we{F}$ depends on $f(\we{r},\we{p},t)$. When the interparticle interactions are given by the potential forces, then $\we{F}(\we{r},t)=-\int \dd{}\we{p'} \dd{}\we{r'}\ \nabla V(\we{r},\we{r'})f(\we{r'},\we{p'},t)$, and thus (1) becomes a self-consistent, nonlinear equation of one particle distribution function. 

Vlasow pointed out that the collision term --- the conventional right-hand side of the Boltzmann equation --- is divergent for long-range Coulomb interactions between charged particles. A replacement of the collision term for a charged particle was suggested by Landau, and its formal derivation was proposed by Rosenbluth [2] in the first glory years of thermonuclear reaction physics. In fact, (1) with the Landau collision term becomes a formidable \m{nonlinear} equation, which plays a fundamental role in plasma physics~[3]. 

Forty years later, the posthumously published book by Vlasov~[4] contained several, mostly failed, attempts to generalize his original ideas, including those for relativistic statistical mechanics. The Vlasov equation has been also used in many condensed matter applications far from classical plasma physics, for example, to describe quark--gluon plasma and some problems in heavy ion collisions~[5] and in the development of late stages of phase separation in the first-order phase transformations~[6]. 

Almost twenty years later, Braun and Hepp~[7] showed that the Vlasov equation describes an asymptotically exact, equal time evolution for the $N$-particle Born, Bogoliubov, Green, Kirkwood, and Yvon (BBBKY) hierarchy with interactions of the form $\tfrac{1}{N}\sum V(\we{r},\we{r'})$. The eigenfunctions of the linearized (1) appear as approximate eigenfunctions of the classical Liouville equation in the Zwanzig variational principle for fluids~[8]. 

Independently from Vlasov, and years later, Yu.L.~Klimontovich~[9,~10] observed that for the \m{$N$-particle} system the ``phase space operator'' 
\beq 
\hat{f}(\we{r},\we{p},t)=\sum\nolimits_{i=1}^N\delta(\we{r}-\we{r_i})\, \delta(\we{p}-\we{p}_i)
\eeq{2}\\[0pt] 
obeys the equation
\beq 
\begin{aligned} 
\aquad \ \frac{\partial \hat{f}(\we{r},\we{p},t)}{\partial t}{+}\frac{\we{p}}{m}\cdot\nabla \hat{f}(\we{r},\we{p},t)
{+}\we{F}\{\hat{f}\}\cdot\frac{\partial \hat{f}(\we{r},\we{p},t)}{\partial \we{p}}{=}0,
\end{aligned} \eeq{}\tyla\\[-13pt]\beq \eeq{3} \tylx \pagebreak 

\noindent where 
\beq 
\we{F}(\hat{f})=\int \dd{}\we{r'}\dd{}\we{p'}\ \we{F}(\we{r},\we{r'})\hat{f}(\we{r'}\we{p'}),
\eeq{4} 
here $\we{F}(\we{r},\we{r'})$ is the force acting between the particles. The fact that the function $\hat{f}(\we{r},\we{p})$ is one of the exact solutions of~(1) plays a fundamental role in all applications of the Vlasov equation and particularly in mathematically correct solutions of it~[3].  

For an $N$-particle system with the Hamiltonian $H(\we{r},\we{p})$, the Hamilton equations of motion, \m{identical} to the Newton ones, can be written as \pagebreak
\beq 
\begin{aligned}
\dot{\we{r}}=\{\we{r},H\},\\[5pt] 
\dot{\we{p}}=\{\we{p},H\},
\end{aligned}
\eeq{}\tyl\beq \eeq{5}
where $\{a,b\}$ denote the Poisson brackets between the arbitrary phase space functions $a=a(\we{r},\we{p})$ and $b=b(\we{r},\we{p})$ 
\beq 
\{a,b\}=\frac{\partial a}{\partial \we{r}}\cdot\frac{\partial b}{\partial \we{p}}-\frac{\partial b}{\partial \we{r}}\cdot\frac{\partial a}{\partial \we{p}}.
\eeq{6}\\[3pt]
Using (5)--(6), we can derive the Poisson bracket relation between the Klimontovich distributions (2),\vs*{3pt} 
\begin{widetext} \tyla
\beq\aquad 
\begin{aligned}
&\Big\{\hat{f}(\we{r},\we{p},t),\hat{f}(\we{r'},\we{p'},t)\Big\}=\Big[\hat{f}(\we{r},\we{p'},t){-}\hat{f}(\we{r'},\we{p},t)\Big]\,{\nabla}\cdot {\nabla}_P\, \delta(\we{r}{-}\we{r'})\, \delta(\we{p}{-}\we{p'})=\hs*{-1mm} \int\hs*{-1mm} \dd{}\we{r''}\dd{}\we{p''}\ \mathcal{C}^{\we{r''}\we{p''}}_{\we{r}\we{p},\we{r'}\we{p'}}
\hat{f}(\we{r''},\we{p''},t), \\
\end{aligned}\eeq{}\tyla
\beq 
\eeq{7}\tyla
\end{widetext} 
\noindent where ${\nabla}\equiv \partial/\partial \we{r}$ and $\nabla_P\equiv \partial/\partial \we{p}$. Now, (7) shows that the algebra of the Klimontovich distribution functions forms the Lie algebra with structure coefficients $\mathcal{C}^{\we{r''}\we{p''}}_{\we{r}\we{p},\we{r'}\we{p'}} $. This algebra is, therefore, of fundamental interest in the metriplectic formulation of dissipative systems dynamics~[11]. 

Assuming the conventional form of the Hamiltonian\tylx
\beq 
 H(\we{r},\we{p})=\sum_{i=1\ldots N}\frac{\we{p}^2}{2m}+\frac{1}{2}\sum_{i<j}V\big(|\we{r}_i-\we{r}_j|\big),
\eeq{8}\\[0pt] 
we can write the Hamiltonian $H$ as functional of $\hat{f}$,\beq 
\aquad\ \begin{aligned}
&H\{\hat{f}\}=\int \dd{}\we{r} \dd{}\we{p}\ \frac{\we{p}^2}{2m}\, \hat{f}(\we{r},\we{p})\\[5pt]
&\quad {+ \frac{1}{2}\int\hs*{-1mm} \dd{}\we{r}\dd{}\we{p}\hs*{-1mm}\int\hs*{-1mm} \dd{}\we{r'}\dd{}\we{p'}\ \hat{f}(\we{r},\we{p})V(|\we{r}-\we{r'}|)\hat{f}(\we{r'},\we{p'})},
\end{aligned}\eeq{}\tyl
\beq
\eeq{9}\tylx
and subsequently (3) can be written as 
\beq 
\frac{\partial \hat{f}(\we{r},\we{p})}{\partial t}=\big\{\hat{f}(\we{r},\we{p}),H\{\hat{f}\}\big\}.
\eeq{10}\\[2pt] 
Note that (2), (7), and (10) form the symplectic formulation of the many-body dynamics equivalent to the Hamiltonian formulation. The complete introduction to symplectic dynamics can be found in a classical book by Marsden and collaborators~[12] and in a series of works by Morrison~[13]. It is worth noticing that by replacing the Klimontovich distribution with the Wigner function~[14] and its Poisson brackets through the Moyal brackets~[15], we obtain the quantum version of the Vlasov--Klimontovich formulation of the many-body system. 

Back in the late 1970s, Piotr Goldstein and~I were working extensively on the use of the Vlasov--Klimontovich formulation to describe the properties of waves propagating in a quasi-relativistic plasma, described by an approximation, in which interactions between charged particles are described by means of the Breit--Darwin Hamiltonian containing velocity-dependent interactions~[16]. \m{Simultaneously,} with Zbigniew Iwi{\'{n}}ski, a former student of Iwo Bia\l{}ynicki-Birula, we were \m{analyzing} the possibility of formulating a fully relativistic form of the Vlasov equation. In a preliminary paper, we formulated such a description and derived the Poisson brackets for a relativistic generalization of the Klimontovich function~[17]. Years later, Iwo Bia\l{}ynicki-Birula and John C. Hubbard were working on the same subject, and eventually, we published together a complete description of the gauge-independent and canonical formulation of the relativistic plasma theory~[18]. \vs*{3pt}

\section{Relativistic plasma theory\vs*{3pt}}

The publication~[18] mentioned at the end of the previous section contained the gauge-independent formulation of the theory of relativistic plasma constituting the multicomponent particle system and the electromagnetic field. Our theoretical tool for that purpose is the symplectic (or canonical) formulation with the dynamical variables for electromagnetic field $\we{E}, \we{B}$ and particle variables, namely, positions $\we{\xi}_A$ and relativistic kinematic momenta $\we{P}_A$. The index $A$-labels particles belonging to a particular particle species $A\in S_a, a=1,\ldots, \mathcal{S}$, and $\we{P}_A=m_a\we{v}_A/\sqrt{1-\we{v}_A^2/c^2}$. These variables obey the Maxwell--Lorentz equations of the form 
\beq
\hs*{-2mm}\begin{aligned}
&\frac{\dd{}\we{\xi}_A}{\dd{}t}=\we{v}_A ,\\[2pt]
&\frac{\dd{}\we{p}_A}{\dd{}t}= e_a\big[\we{E}\big(\we{\xi}_A(t),t\big)+\we{v}_A(t)\times\we{B}\big(\we{\xi}_A(t),t\big)\big], \\[2pt]
&\frac{\partial \we{B}(\we{r},t)}{\partial t}=-{\nabla}\times\we{E}(\we{r},t), \\[2pt]
&\frac{\partial \we{E}(\we{r},t)}{\partial t}={\nabla}\times\we{B}(\we{r},t)-\hs*{-1mm}\sum_A e_A\we{v}_A(t)\,\delta(\we{r}{-}\we{\xi}_A (t)),\\[2pt]
&{\nabla}\cdot\we{B}(\we{r},t)=0, \\[2pt]
&\nabla\cdot \we{E}(\we{r},t)=\sum_A e_A\, \delta(\we{r}-\we{\xi}_A (t)).
\end{aligned}\eeq{}\tyl \tyla\beq \eeq{11} 
\noindent The relativistic invariant phase space Klimontovich function is identical to that in (2), with the \m{relativistic} kinematic momenta replacing $\we{p}$, i.e., \tylx\pagebreak
\beq 
\hat{f}_a(\we{r},\we{p},t)=\sum\nolimits_{A\in\mathcal{S}_a}\delta(\we{r}-\we{\xi}_A)\delta(\we{p}{-}\we{P}_A),
\eeq{12}\\[2pt]
where 
\beq 
\we{p}={m_a\we{v}}/{\sqrt{1-\we{v}^2/c^2}}.
\eeq{13}\\[4pt]
\indent The Maxwell--Lorentz equation (11) can be cast into a canonical form using the Poisson bracket relations
\beq \begin{aligned}
&\{ \xi^i_A,P^j_B\}=\delta_{AB}\, \delta^{ij},\\[5pt]
&\{P^i_A,P_B^j\}=e_A\delta_{AB}\, \epsilon^{ijk} B^k(\we{\xi}_A), \\[5pt]
& \{ P^i_A,E^j(\we{r})\}=e_A\delta^{ij}\, \delta(\we{r}-\we{\xi}_A),\\[5pt]
&\{B^i(\we{r}),E^j(\we{r'})\}=\epsilon^{ijk}\partial_k\, \delta(\we{r}-\we{r'}),\\
\end{aligned}\eeq{} \tylx\tyla \beq \eeq{14}\pagebreak 
with all other \m{Poisson} brackets vanishing. The Poisson brackets for electromagnetic fields $\{E^i,B^j\}$ are the classical form of commutators derived by Born and Infeld~[19] and discussed in greater detail in~[20]. These Poisson brackets are consistent with the constraints described by the last two equations in~(11).\\
\indent With the above choice of canonical variables and their Poisson brackets, the full Poincar\'{e} group is realised as a subgroup of the canonical transformation group~[18] and the theory of plasma becomes fully relativistic. Using (14) one can easily derive the Poisson brackets for electromagnetic fields and the phase space function (12) which we write below employing shorthand notation $\we{z}{=}(\we{r},\we{P})$, $\we{\zeta}_A{=}(\we{\xi}_A,\we{P}_A)$, rational system of units with $c=1$ and following\vs*{0pt}
\beq 
\aquad\ \Big\{
F\big(\{\we{\zeta}_A\}\big), G\big(\{\we{\zeta}_B\}\big)\Big\}=\sum_{\zeta_A,\zeta_b}\frac{\partial F}{\partial \zeta_A}\Big\{\zeta_a,\zeta_B\Big\}\frac{\partial G}{\partial \zeta_B}, 
\eeq{}\tyla
\beq
\eeq{15}
\begin{widetext}  \tyla
\beq 
\big\{\hat{ f}_a(\we{z},)\hat{f}_b(\we{z'})\big\}=
\delta_{ab}\Big[\Big(\hat{f}_a(\we{r},\we{P'})-\hat{f}_a(\we{r'},\we{P})\Big){\nabla}\cdot{\nabla}_{P}
+e_a \we{B}(\we{r})\cdot \big({\nabla}_P \hat{f}_b(\we{z})\times{\nabla}_P\big)\Big]\delta(\we{z}-\we{z'}).
\eeq{16}\\[3pt]
\indent The remaining non-zero Poisson brackets read
\beq
\big\{\hat{ f}_a(\we{z}),\we{E}(\we{r'})\big\}=-e_a{\nabla}_P \hat{ f}_a(\we{z})\delta(\we{z}-\we{z'}),
\quad\big\{\hat{ f}_a(\we{z}),\we{B}(\we{r'})\big\}=0.
\eeq{17} \tylx
\end{widetext}
Having the above formalism, we can express all the generators of the Poincar\'{e} group in terms of the Klimontovich function~$\hat{f}$ and fields $\we{E}$, $\we{B}$~[18]. For example, the Hamiltonian of the system reads
\beq 
\aquad \ {H}=\sum_a\int\hs*{-1mm} \dd{}\we{z}\ \sqrt{\we{p}^2{+}m^2}\hat{f}_a(\we{z})+\frac{1}{2}\int\hs*{-1mm} \dd{}\we{r}\, \big(\we{E}^2{+}\we{B}^2\big),
\eeq{}\tyl\beq \eeq{18}\\[1pt]
and momentum vector
\beq 
\we{\Pip}=\sum_a\int \dd{}\we{z}\ \we{p}\,\hat f_a(\we{z})+\int \dd{}\we{r}\ \we{E}\times\we{B} .
\eeq{19}\\[2pt]
Note that the Hamiltonian (18) does not contain the coupling constant between the plasma and the electromagnetic field, i.e., the charge $e_a$. The interaction between these two is fully contained in the Poisson brackets (14), (16), (17). The Klimontovich--Vlasov formulation of relativistic plasma physics, presented in~[18], therefore follows some ideas presented by Souriau and Sternberg~[21,~22]. This gauge-invariant formulation of the interacting system of particles and fields can be extended for general relativity formulation~[23]. 

The Vlasov equation and the Maxwell equation can then be written as
\beq \begin{aligned}
&\frac{\partial \hat{f}_a}{\partial t}= \Big\{\hat{f}_a,H\big\{\hat{f}_a,\we{E},\we{B}\big\}\Big\}\equiv \\
& \quad 
 - \Big[\we{v}_a\cdot{\nabla}+e_a \big(\we{E}+\we{v}_a\times\we{B}\big)\cdot{\nabla}_P\Big]\hat{f}_a, \\
& \frac{\partial \we{E}}{\partial t} =\{\we{E},H\}\equiv {\nabla}\times\we{B}-\sum\nolimits_a\int\nolimits \hs*{-1mm}\dd{}\we{p}\ \we{v}_a\hat{f}_a, \\
& \frac{\partial \we{B}}{\partial t}=\{\we{B},H\}\equiv-{\nabla}\times\we{E}.
\end{aligned}\eeq{}\tyl\beq \eeq{20}\tyla \pagebreak

The relativistic statistical mechanics does not offer a mathematically rigorous formulation of the relation between the Vlasov (1) and the Kimontovich equation (3) like that in~[7]. Nevertheless, there is sufficient experimental experience from hot plasma and astrophysical applications for one to make an assumption that the one-particle distribution function defined in one-particle phase space of positions and relativistic kinematical momenta --- the ensemble average of the Klimontovich function --- obeys identical equations as our relativistic one (20). With this assumption, we can generalize the formulation given above by including in our description direct information on destroying physical processes in plasma --- direct charge particle collisions --- similarly as Landau has done for the original Vlasov equation. To do this, it is convenient to follow an algebraic method of including dissipative processes in symplectic dynamics --- the metriplectic method. We shall discuss this procedure in Sect.~4.\vs*{4pt}

\section{Semiclassical spin 1/2 Bloch\\electrons plasma\vs*{4pt}}
 
In the quantum theory of crystalline solids, the motion of electrons is described by means of the wave packets constructed from Bloch wave functions with periodic part $u_{\we{k}}$, where $\we{k}$ labels the wave vectors for the specific band~[24]. For the sake of simplicity, we consider here the solids with only one energy band $\epsilon(\we{k})$. J.~Zak~[25] observed that the Bloch systems yield the geometric phases and that the gauge-invariant Berry curvature~[26] \pagebreak
\beq 
\bm{\mathit{\widetilde{\Omega}}}(\we{k})= \ii \big\langle {\nabla}_{\we{k}} u_{\we{k}}|\times|{\nabla}_{\we{k}}u_{\we{k}}\big\rangle
\eeq{21}\\[2pt]
is observable and generally nonzero for crystals without inversion symmetry. Theoretical analysis~[27,~28] has shown that in many important experimental applications, it is sufficient to describe the motion of electrons by a semiclassical equation of motion in which the position of the center of the localized electron wave function and wave vector ($\we{r},\we{k}$) obey the equation of motion
\beq
\dot{\we{r}}=\frac{\partial \epsilon(\we{k})}{\hbar\, \partial \we{k}}+\we{F}\times\bm{\mathit{\widetilde{\Omega}}}/\hbar,
\quad \dot{\we{k}}=\we{F}/\hbar,
\eeq{22}\\[2pt]
where $\we{F}=-{\nabla}U(\we{r})$ is the net force acting on the electrons. 

These equations play the role of the Hamilton equations in classical mechanics and, therefore, can be used in the formulation of the symplectic, or Lie--Poisson bracket technique~[11,~12,~29], description of the semiclassical Bloch electrons. This description can subsequently be rewritten using the Vlasov--Klimontovich equation approach~[30]. We begin by defining the Poisson brackets for ``position'' $\we{r}$ and ``momenta'' $\we{\kappa}=\hbar\we{k}$ (where $\we{\Omegap}  =\bm{\mathit{\widetilde{\Omega}}}  /\hbar)$ as follows
\beq 
\begin{aligned}
& \{r_a,r_b\}=\epsilon_{abc}\Omegap _c, & \{r_a,\kappa_b\}=\delta_{ab},\\[4pt]
& \{\kappa_a,\kappa_b\}=0, & \\
\end{aligned}\eeq{}\tyl
\beq\eeq{23}
a special case of non-commutative classical mechanics Poisson brackets discussed in~[31]. Assuming that the Hamiltonian for semiclassical electrons can be written as $H(\we{r},\we{\kappa})=\epsilon(\we{\kappa})+U(\we{r})$ equations
\beq 
\dot{r}_{a}=\{r_a,H\},\qquad \dot{\kappa}_a=\{\kappa_a, H\},
\eeq{24} 
become identical to~(22). That allows us to use the Klimontovich function~[9,~10] with momenta $\we{p}$ replaced by $\we{\kappa}$ to describe the semiclassical Bloch electron plasma. We can write, as in previous sections, $H\{\hat{f}\}=\int \dd{}\mathbf{1}\, \epsilon(\mathbf{1})\hat{f}(\mathbf{1})$, where $\mathbf{1}=(\we{r},\we{\kappa})$ and the mean value of a physical observable as $\langle A \rangle =\int\hs*{-1mm} \dd{}\mathbf{1}\, A(\mathbf{1})\hat{f}(\mathbf{1})$. Using the Poisson brackets (23), one easily finds the Poisson brackets for $\hat{f}$ function
\beq 
\begin{aligned}
&\hs*{-3mm} \big\{\hat{f}(\mathbf{1}),{\hat{f}}(\mathbf{2})\big\}=
\Big({\nabla}\hat{f}(\mathbf{1})\cdot \nabla_{\we{\kappa}} {-} {\nabla}_{\we{\kappa}}\hat{f}(\mathbf{1})\cdot{\nabla}\Big)\delta(\mathbf{1}{-}\mathbf{2})\\[2pt]
&\quad  +\we{\Omegap}  \cdot\Big[{\nabla}\hat{f}(\mathbf{1})\times{\nabla}\delta(\mathbf{1}{-}\mathbf{2}))\Big].
\end{aligned}\eeq{}\tyl \tyl\\[-4pt]\beq 
\eeq{25}\tyla

Now, (25) allows us to write the Vlasov--Klimontovich equation for the semiclassical electrons as 
\beq 
\frac{\partial \hat{f}(\mathbf{1})}{\partial t}=\{\hat{f}(\mathbf{1}), H\{\hat{f}\}\}.
\eeq{26}\\[4pt] 
\indent To account for the specific properties of the semiclassical electron plasma, for example, Ohm's law, we need to supplement the Vlasov--Klimontovich equation (26) with the proper dissipative term on its RHS, which we shall denote $W\{\hat{f}\}$. In~[27,~28], the simple relaxation time approximation has been used for $W\{\hat{f}\}$. The full kinetic equation for $\hat{f}$ does not then conserve the number of charge carriers in the system. It seems, therefore, more appropriate to replace this $W\{\hat{f}\}$ by the generalization of the Boltzmann--Lorentz collision operator~[32,~33], which offers a formulation of the collision operator for a~tight-binding model. The relation between the construction of such an operator and the symplectic formulation of many-particle system dynamics will be outlined in the last section. Having done so and using the linearized version of the Chapman--Enskog approximation $\hat{f} (\we{r},\we{\kappa})\approx\rho(\we{r})\phi_B(\we{\kappa})$, where $\rho(\we{r})$ denotes carriers density and $\phi_B(\we{\kappa})$ stands for equilibrium carriers distribution function at the temperature of the carrier $\beta^{-1}$ defined as $m\delta_{ij}\beta^{-1}=\int \dd{}\we{\kappa}\ \phi_B(\we{\kappa})\kappa_i\kappa_j$, and, furthermore, assuming that the Berry curvature $\we{\Omegap}$ is slowly varying function of the wave vector $\we{k}$ traversing the Brillouin zone, we obtain the dispersion relation for the fluctuation of the density of carriers $\rho_{\omega,\we{q}}$, which replaces the Ohm law for spinless Bloch electrons
\beq \begin{aligned}
&\Big(\tilde{\omega}(\we{q})+\ii  z \Gammap_{\we{q}} \Big) \Big(\tilde{\omega} (\we{q})-\ii  S_{\we{q}} \Gammap_{\we{q}}\Big)= \\
&\quad \Big[ \frac{\we{q}^2}{m\beta}-e\we{q}\cdot \Big(\we{E}_0\times(\we{q}\cdot \overline{\we{\Xip}})\Big)+\ii  e\we{E}_0\cdot \frac{\we{q}}{m}\Big],
\end{aligned}
\eeq{}\tyl\beq\eeq{27}\\[3pt]
where 
\beq 
\tilde{\omega}(\we{q})=\omega -e \we{q} \cdot (\overline{\we{\Omegap}}\times\we{E}_0)
\eeq{28}\\[3pt]
is the frequency shift due to the anomalous Hall drift velocity $e\overline{\we{\Omegap}  }\times \we{E}_0$. There, $\Gammap_{\we{q}}$, $S_{\we{q}}$, and $z$ are the scattering amplitude, scatterers structure factor, and coordination number, respectively; $\overline{\Omegap}$~and~$\overline{\Xi}_{ij}$ denote, respectively, the averaged values of the Berry curvature $\overline{\Omegap }=\int \dd{}\we{\kappa}\, \phi_B( \we{\kappa})\we{\Omegap}(\we{\kappa})$ and the averaged ``curvature torque'' $\overline{\Xip}_{ij}=\int \dd{}\we{\kappa}\, \kappa_i\,\Omegap_j(\we{\kappa})\phi_B(\we{\kappa})$.

The charge carriers in solid carry internal degrees of freedom spins. One can generalize the Vlasov--Klimontovich description of semiclassical carriers, shown above, for the case of Bloch electrons with spin $\tfrac{1}{2}$. We do this by describing the carriers by spinor Klimontovich distribution function \m{$\hat{f}=\frac{1}{2}\sum_{\alpha=0}^3 f_\alpha \hat{\sigma}_\alpha$,} where $\hat{\sigma}_{i=1,2,3}$ are the Pauli matrices and $\sigma_0$ is the $2\times 2$ unit matrix. The meaning of coefficients $f_{\alpha}$ stems from the meaning of the mean value of the observables $\langle{}A\rangle=\Tr(A\,\hat{f})$. Here, $\Tr$ denotes the matrix trace in spinor space and phase-space integration. The coefficient $f_0$ is the Vlasov--Klimontovich function used for spinless carriers, and $f_i$ are carriers spin densities $\langle S_i\rangle=\frac{1}{2}\int \dd{}\mathbf{1}\, \Tr(\frac{\hbar}{2}\hat{\sigma}_i\sum_\alpha f_\alpha \hat{\sigma}_\alpha)=\frac{\hbar}{2}\int \dd{}\mathbf{1}\, f_i(\mathbf{1})$.

For the spinor distribution function $\hat{f}$, the Poisson brackets now become $4\times 4$ functional matrices ($i,j=1,2,3$)
\beq 
\aquad  \left[ \hat{f}({\bf{1}}),\hat{f}(\bf{2}) \right] {=} \left(
{\begin{array}{cc}
  {\left\{ {f_0 ({\bf{1}}),f_0 ({\bf{2}})} \right\},} & {\left\{ {f_0 ({\bf{1}}),f_j ({\bf{2}})} \right\}}\\[3pt]
  {\left\{ {f_i ({\bf{1}}),f_0 ({\bf{2}})} \right\},} & {\left\{ {f_i ({\bf{1}}),f_j ({\bf{2}})} \right\}}\\
\end{array}} \right),  
\eeq{}\tyla \beq \eeq{29}\\[2pt]
where $\{\hat{f}_0(\mathbf{1}),\hat{f}_0(\mathbf{2})\}$\} is given by (25) and
\beq \begin{aligned}
& \big\{\hat{f}_i(\mathbf{1}),\hat{f}_0(\mathbf{2})\big\}=-{\nabla}\hat{f}_i(\mathbf{1})\cdot{\nabla}_{\we{\kappa}}\, \delta(\mathbf{1}{-}\mathbf{1})\nonumber\\[3pt]
& \big\{\hat{f}_i(\mathbf{1}),\hat{f}_j(\mathbf{2})\big\}=\epsilon_{ijk}\hat{f}_k(\mathbf{1})\, \delta(\mathbf{1}{-}\mathbf{1}).
\end{aligned}\eeq{}\tyl \beq \eeq{30}\tyla\pagebreak

The kinetic equation for the spinor $\hat{f}$ becomes now\beq \begin{aligned}
&\frac{\partial\, \hat{f}}{\partial t}=\big[\,{\hat{f},H\{\hat{f}\}}\big] +{\cal W}\{ \hat f \} =\\
\end{aligned}\eeq{}\tyla
\beq\quad \Tr_\mathbf{2}\big[\, \hat{f}(\mathbf{1}), \hat{f}(\mathbf{2})\big]\, \frac{\delta H\{ \hat{f} \}}{\delta \hat{f} (\mathbf{2})}+{\cal W}\{ \hat{f} \}. \eeq{31}\\[3pt]
The collision operator ${\cal W}$ must preserve the length of the carrier spin $\we{S}^2$. The Poison brackets for the spinor $\hat{f}$ (see~(29)) guarantee that the spin length is the Casimir of the Lie--Poisson brackets.  

Using the notation from~[34,~35], we write 
\beq 
\hs*{-3mm}{W}\{ \hat{f} \}=
\eeq{}\tyla\tylx\beq 
{-}\lambda \int\hs*{-1.5mm}\dd{}\mathbf{2} \Big[
\delta_{ij}\phi_B(\mathbf{\kappa}_2) \delta(\mathbf{1}{-}\mathbf{2})-\hat{f}_i(\mathbf{1})\hat{f}_j(\mathbf{2})\Big]\frac{\delta H\{\hat{f} \}}{\delta\hat{f}(\mathbf{2})}.
\eeq{}
\tylx\beq \eeq{32}
Having formulated the kinetic equation for semiclassical spin $1/2$ Bloch electrons, we can use them to analyze the properties of such a carrier plasma. For example, we can consider plasma with constant equilibrium carriers density $\rho_0$, constant external electric field $\we{E}_0$, and diagonal pressure tensor $P_{ij}(\rho)=\int \dd{}^3\kappa\ f(\we{\kappa})\kappa_i\kappa_j$. Linearizing (31), we~obtain the dispersion relation for plasma excitation\beq 
\tilde{\omega}(\we{q})^2=\omega_P^2+C_{ij}\, q^iq^j+\frac{\ii  e}m{}\ \we{q}\cdot\we{E}_0,
\eeq{33}\\[3pt] 
where $(\omega, \we{q})$ denotes the plasma excitations frequency and momentum; $\omega_P{=}\sqrt{4\pi e^2\rho_0 /m}$ is the plasma frequency; $\tilde{\omega}(\we{q})$ is given in (28); $c{=}(\frac{\partial P}{\partial \rho})^{1/2}_{_{0}}$ is the speed of sound in carrier gas at equilibrium; and sound velocity is anisotropic $c\rightarrow C_{ij}q^iq^j$, with $C_{ij}=c^2(\delta_{ij}{-}\epsilon_{ijk}\,E_0^l \,\overline{\Xip}^{kl}/c^2 )$. This anisotropy is caused by the coupling of the external electric field and the Berry phase curvature torque. The plasma excitations group velocity is shifted with respect to that of usual plasma in the frame of reference drifting with the anomalous Hall velocity \m{$\we{v}_D=e \we{\widetilde{\Omegap } }\times\we{E}_0$.} 
 
In the following sections, we shall discuss the metriplectic generalization of the Vlasov--Klimontovich formulation of both relativistic plasma and plasma of semiclassical Bloch electrons.\vs*{5pt}

\section{Metriplectic description\vs*{4pt}}

In two previous sections, we have described the use of the Lie--Poisson brackets technique, together with the use of the Vlasov--Klimontovich function, for two important physical models of plasmas: the relativistic plasma and the plasma of semiclassical Bloch, spinless and spin 1/2, electrons. The Lie--Poisson bracket technique has been applied to many other examples in non-linear physics leading to important progress in those fields~[12,~36]. All these applications are examples of reversible (dissipationless) dynamical systems. Allan Kaufman and Phil Morrison suggested~[\m{37--39}] that this description can be generalized to include dissipative processes by employing a technique called now metriplectic dynamics. 
    
This theory, described for example in~[11,~13], consists of two steps. 

Step one is replacing the Hamiltonian in equations of motions by the system free energy $\cal{F}(\psi)=H(\psi){-}\theta\, S(\cal{C})$, where $\psi$ stands for the system \m{dynamical} variables, $S$ is the entropy functional~and\linebreak $\cal{C}$ denotes a set of the Casimir variables defined as quantities which Poisson brackets with all $\psi$ vanishes identically, independently of the form of the Hamiltonian~[29]. The coefficient $\theta$ depends on the type of interaction between the dynamical system and the environment. For example, it can be identified with the system temperature by assuming that the absolute minimum of $\cal{F}(\psi)$ is described~by the equilibrium distribution function for the system given by the Hamiltonian~$H$. In both examples discussed in previous sections, this distribution function is the proper Maxwell--Blotzmann distribution. 

Step two consists in adding to the Poisson brackets in the Hamilton equations of motion (5) the symmetric--``dissipative'' brackets
 \beq 
\aquad \ {\prec{\hs*{-1mm}\psi(\zeta_A),\psi(\zeta_B)}\hs*{-1mm}\succ}={\prec{\hs*{-1mm}\psi(\zeta_B),\psi(\zeta_A)}\hs*{-1mm}\succ}=D(\psi_A,\psi_B).
\eeq{}\tyl\beq 
\eeq{34} 
The dynamical equations of motion are now written as
\beq 
\aquad \frac{\partial \zeta_A}{\partial t}=\Big[\psi_A(\zeta),\cal{F}(\zeta) \Big]=\int {\rm D\zeta'}\, \cal{L}_{AB}(\zeta,\zeta')\, \frac{\delta \cal{F}}{\delta \psi_B(\zeta')},
\eeq{}\tyla\beq \eeq{35}\\[-4pt] 
where
\beq 
\aquad\  \cal{L}_{AB}(\zeta,\zeta')=\big\{\psi_A,\psi_B\big\}- {\prec\hs*{-1mm}{\psi(\zeta_A),\psi(\zeta_B)}\hs*{-1mm}\succ}. 
\eeq{36} 
\tyla

The structure of the dissipative bracket depends on the nature of the processes causing the dissipation. 

In~[29], the general theory of algebraic construction of $D(\psi_A,\psi_B)$ was given. For the dynamical theory equipped with Lie--Poisson brackets associated with Lie algebra with the structure constant $C^A_{BC}$ (compare with (7)), the dissipative brackets, consistent with preservation of the Casimirs of that Lie algebra, have the form
\beq 
D_{AB}=G^{CD}C^M_{CA}\,C^N_{DB}\,\psi_M\psi_N,
\eeq{37} 
where $G^{CD}$ is the inverse of the Cartan--Killing tensor built from the structure constant as \m{$G_{AB}=-C^D_{AN}C^N_{DB}$.}

In both applications described in previous sections, these brackets describe the direct particle--particle collisions. Thus recalling that the role of fields $\psi$ is played by the Klimontovich function, we can rewrite (34) as
\beq 
{\prec\hs*{-1mm}{\hat{f}(\we{z}),\hat{f}(\we{z}')}\hs*{-1mm}\succ}\equiv D(\hat{f}(\we{z}),\hat{f}(\we{z}'))=
\eeq{}\tyla
\beq\quad
\frac{1}{2}\int \dd{}^6z_1 \dd{}^6z_2\ \hat{f}(\we{z}_1)\Deltap(\we{z},\we{z}';\we{z}_1,\we{z}_2)\hat{f}(\we{z}_2),
\eeq{} \tyl \beq \eeq{38} 
where the kernel $\Deltap(\we{z},\we{z}';\we{z}_1,\we{z}_2)$ accounts for physics of those particle--particle interactions. 

In the analysis of the semiclassical Bloch electrons, we have used the collision operator $\cal{W}$. The use of a linearized version of such operator results in the dispersion relation in~(27). A similar collision operator is shown in (32). 

The dissipative bracket for relativistic classical plasma, discussed earlier, should yield the collision operator on the RHS of the first equation in (20) in the form of the Landau collision operator~[3]. This requirement gives the kernal $\Deltap(\we{z},\we{z}';\we{z}_1\we{z}_2)$ in the form
\begin{widetext}\tyla
\beq 
\Deltap(\we{z},\we{z}';\we{z}_1\we{z}_2)=\int \dd{}\we{k}\ \alpha_{\we{k}}(\we{z}_1,\we{z}_2)\, \delta\big(\we{k}\cdot(\we{v}(\we{p}){-}\we{v}(\we{p}')\big)\  \left(\we{k}\cdot\Big[{\nabla}_{P_1}\,\delta(\we{z}{-}\we{z}_1)- {\nabla}_{P_2}\,\delta(\we{z}{-}\we{z}_2)\Big]\right)
\eeq{}\tylx \beq 
\qquad \times\ \left(\we{k}\cdot\Big[{\nabla}_{P_1}\, \delta(\we{z}'{-}\we{z}_1)- {\nabla}_{P_2}\, \delta(\we{z}'{-}\we{z}_2)\Big]\right),
\eeq{39}  \tylx
\end{widetext}
where $\alpha_{\we{k}}(\we{z}_1,\we{z}_2)$ describe details of particle--particle collision and as before, $\we{v}(\we{p})=\we{p}/\sqrt{\we{p}^2+m^2}$,~[40].

The metriplectic description based on the Klimontovich function suffers from mathematical difficulties related to the problems with the operations on the singular distribution functions. In most of the applications, the distribution $\hat{f}$ is therefore replaced by the ``smooth'' one-particle distribution function $f(\we{r},\we{p},t)=\langle \hat{f}(\we{r},\we{p},t)\rangle$, where $\langle\ldots\rangle$ denotes initial ensemble averaging. That surely leads to the loss of information. One can attempt to restore at least part of that lost information by amending the RHS of (35) with a properly chosen Langevin ``force''~\m{[40--42] }
\beq 
\frac{\partial \zeta_A}{\partial t}=\int {\rm D \zeta'}\ \cal{L}_{AB}(\zeta,\zeta')\frac{\delta \cal{F}}{\delta \psi_B(\zeta')}+\lambda_A(\zeta)
\eeq{}\tyl\beq \eeq{40} 
with
\beq 
\hs*{-2mm}\langle\langle\lambda_A(\zeta,t)\lambda_B(\zeta',t)\rangle\rangle= S_{AB}(\zeta,\zeta')\, \delta(t-t'),
\eeq{41}\\[3pt]
where the double brackets $\langle\langle\ldots\rangle\rangle$ denote averaging over the realizations of the Langevin forces $\lambda_A$. The generalized Fokker--Planck equation for the probability distribution $\cal{P}$ in space of dynamical variables $\psi$ can now be written as~[40]
\beq 
\frac{\partial \cal{P}}{\partial t}= \hat{L}\left(\frac{\delta}{\delta \zeta}\right)\cal{P},
\eeq{42} 
where
\beq \begin{aligned}
&\cal{P}=\int {\rm D \zeta}\, {{\rm D}\zeta'}\ \frac{\delta}{\delta\psi_A(\zeta)}\cal{L}_{AB}(\zeta,\zeta')\frac{\delta \cal{F}}{\delta \psi_B(\zeta')}\\ 
&\quad +\int {{\rm D} \zeta}\, {{\rm D}\zeta'}\ \frac{\delta}{\delta\psi_A(\zeta)}\cal{S}_{AB}(\zeta,\zeta')\frac{\delta }{\delta \psi_B(\zeta')}.
\end{aligned}\eeq{} \tyl\beq \eeq{43}\tyla

Note that (43) has the same form for both specific examples discussed in Sect.~3. Note, therefore, that for the relativistic plasma, it is a fully relativistic Fokker--Planck equation for the dynamical variables $\psi_A(\zeta)$, which in this case are the Vlasov \m{one-particle} distribution function $f(\we{r},\we{p})$, with $\we{p}$ given by (13) and electromagnetic field ($\we{E},\we{B})$. For semiclassical Bloch electrons, $\we{p}$ is the kinematic momentum of the carrier. Assuming that the Hamiltonian for semiclassical Bloch carriers is given as a~nonrelativistic form of (18) supplemented with the Zeeman-like coupling proportional $\sum_{j=1}^3 \gamma\, f_jB_j(\mathbf{1})$ and neglecting the internal electric and magnetic fields generated by the motion of carriers and using the dissipative brackets (32) stemming from dissipative brackets for spins~[43]
\beq 
\prec S_i,S_j\succ=-\lambda |\we{S}|\, \Big(\delta_{ij}-\frac{S_i S_j }{\we{S}^2}\Big),
\eeq{44}\\[2pt]
we can derive the conservation equation for the spin density $S_i(\we{r})=\int \dd{}\we{p}\, f_i(\we{r},\we{p})$, which is equivalent to the convective version of the Gilber--Landau equation for sample magnetization~[44]
\beq 
\frac{\partial \we{S}}{\partial t}+{\nabla}\cdot \we{\Lambdap}=\gamma\we{S}\times\we{B}-\lambda\, \we{S}\times \big(\we{S}\times\we{B}\big),
\eeq{45} 
where $\Lambdap_{ij}=\int \dd{}\we{p}\ p_i\, f_j(\we{r},\we{p})$ is the spin current tensor. The explicit form of $\we{\Lambdap}$ follows from the Chapman--Enskog approximations in solving the kinetic equation (31).

The above example of continuum equations following from metriplectic analysis of the Vlasov--Klimontovich description of the many particles system allows us to derive the hydrodynamic-like description of that system. These continuum mechanics equations can also be cast in the form of metriplectic dynamics~[41], and it has been discussed in many recent publications of Massimo Materassi and Phil Morrison and their collaborators~[13,~36]. Whether this technique can be useful in other applications, for example, in the theory of quark--gluon plasma, remains to be seen.

\vs*{5pt}
\section{Conclusions\vs*{5pt}}

The above sections contain a discussion of the use of Vlasov--Klimontovich formulation of the classical many particle systems dynamics. Some important generalisations of that formulation, for {example,} quantum many-body problems or {classical} hydrodynamics~[36] are mentioned in included references. There is essentially no applications of that formulation in equilibrium statistical mechanics in spite of the fact that the Vlasov description could easily be used within the Martin--Rose--Sigma formulation~[45]. The general {relativity} {generalisation} of the kinetic theory base on the present above \m{formulation~[23]} is now being prepared for \m{publication.}  

\AC \vs*{6pt}

This mini-review is based on collaborative work with many colleagues and students whose contributions are acknowledged in the paper bibliography.

Iwo Bia\l{}ynicki-Birula was the mentor for many generations of physicists, in particular, to the group which joined him at the Center for Theoretical Physics in 1979 and continues to work at the same institution until today. I personally benefited throughout all that time from intellectual interactions with Professor Iwo Bia\l{}ynicki-Birula and from his support and friendship in the bright and dark periods of that institution and the Country. We hope to continue to benefit from his remarkable wisdom and confidence for a better future of science and the world for many more years.

\mbox{}\\[5pt]
{\bf References}
\mbox{}\\[5pt]
\begin{enumerate}
\item{A.A. Vlasov, \refdo{{\em J. Exp. Theor. Phys.} {\bf 3}, 291 (1938)}{} (in Russian)}

\item{M.N. Rosenbluth, W.M. MacDonald, D.L. Judd \refdo{{\em Phys. Rev.} {\bf 107}, 1 (1957)}{10.1103/PhysRev.107.1}}

\item{N.G. van Kampen, B.U. Felderhof, \refdo{\rm Theoretical Methods of Plasma Physics}{}, Cambridge University Press, 2009}

\item{A.A. Vlasov, \refdo{\em Nonlocal Statistical Mechanics}{}, Nauka, Moskow 1978 (in Russian)}

\item{\refdo{\em Quark-Gluon Plasma: Theoretical Foundations: An Annotated Reprint Collection}{}, Eds. J. Kapusta, B. M\"uller, J. Rafelski, 2003}

\item{G. Manzi, R. Marra, in: \refdo{\em Transport Phenomena and Kinetic Theory, Applications to Gases, Semiconductors, Photons, and Biological Systems}{}, Eds. C. Cercignani, E.~Gabetta, Birkh\"aser, 2007}

\item{W. Braun, K. Hepp, \refdo{{\em Commun. Math. Phys.} {\bf 56}, 101 (1977)}{10.1007/BF01611497}}

\item{R. Zwanzig, \refdo{{\em Phys. Rev.} {\bf 144}, 170 (1966)}{10.1103/PhysRev.144.170}}

\item{Y.L. Klimontovich, \refdo{{\em Zh. Eksp. Teor. Fiz.} {\bf 34}, 173 (1958)}{}}

\item{Y.L. Klimontovich, \refdo{\em Statistical Theory of Nonequilibrium Processes in Plasma}{}, MIT Press, Cambridge 1984}

\item{L.A. Turski, in: \refdo{\em Continuum Models and Discrete Systems}{}, Vol. 1, Ed. G.A. Maugin, Longman, London 1990}

\item{J.E. Marsden, T.S. Ratiu, \refdo{\em Introduction to Mechanics and Symmetry}{}, Springer-Verlag, Berlin 2002}

\item{P.J. Morrison \refdo{{\em J. Phys. Conf. Ser.} {\bf 169}, 012006 (2009)}{10.1088/1742-6596/169/1/012006}}

\item{E.P. Wigner, \refdo{{\em Phys. Rev.} {\bf 40}, 749 (1932)}{10.1103/PhysRev.40.749}}

\item{J.E. Moyal, \refdo{{\em Proc. Cambr. Phil. Soc.} {\bf 45}, 99 (1949)}{10.1017/S0305004100000487}}

\item{P. Goldstein, L.A. Turski, \refdo{{\em Physica A} {\bf 89}, 481 (1977)}{10.1016/0378-4371(77)90077-2}}

\item{Z.R. Iwi\'nski, L.A. Turski, \refdo{{\em Lett. Appl. Sci. Eng.} {\bf 4}, 179 (1976)}{}}

\item{I. Bia\l{}ynicki-Birula, J.C. Hubbard, L.A.~Turski, \refdo{{\em Physica A} {\bf 128}, 509 (1984)}{10.1016/0378-4371(84)90189-4}}

\item{M. Born, L.~Infeld, \refdo{{\em Proc. R. Soc. London}{} {\bf 150}, 141 (1935)}{10.1098/rspa.1935.0093}}

\item{J.E. Marsden, A.~Weinstein, \refdo{{\em Physica D} {\bf 4}, 394, (1982)}{10.1016/0167-2789(82)90043-4}}

\item{J.M. Souriau, \refdo{\em Structure des Syst\'ems Dynamique}{} Dunod, Paris 1970}

\item{J.F. Carine\^na, H.~Figueroa, P.~Guha, \arx{1501.04917v2}, 2015}

\item{I.~Bia\l{}ynicki-Birula, L.A.~Turski, to be publish}

\item{N.W. Ashccrofy, N.D. Mermin, \refdo{\em Solid State Physics}{}, Philadelphia 1976}

\item{J. Zak, \refdo{{\em Phys. Rev. Lett.} {\bf 62}, 2747 (1989)}{10.1103/PhysRevLett.62.2747}}

\item{M.V. Berry, \hreff{www.jstor.org/stable/2397741}{{\em Proc. R. Soc. A} {\bf 392}, 45 (1984)}}

\item{D. Culcer, J. Sinova, N.A. Sinitsyn, T.~Jungwirth, A.H. MacDonald, Q. Niu, \refdo{{\em Phys. Rev. Lett.} {\bf 93}, 046602 (2004)}{10.1103/PhysRevLett.93.046602}}

\item{G. Sundaram, Q. Niu, \refdo{{\em Phys. Rev. B} {\bf 59}, 14915 (2002)}{10.1103/PhysRevB.59.14915}}

\item{L.A. Turski, \refdo{Dissipative Quantum Mechanics}{}, Springer Lect. Notes in Physics 4777, Eds. Z. Petru, J. Przystawa, K. Rapcewicz, New York 1996}

\item{W.C. Kerr, M.J. Rave, L.A. Turski, \refdo{{\em Phys. Rev. Lett.} {\bf 94} 176403 (2005)}{10.1103/PhysRevLett.94.176403}}

\item{A.E.F. Djemai, \refdo{{\em In. J. Theor. Phys.} {\bf 43}, 299, (2004)}{10.1023/B:IJTP.0000028864.02161.a3}}

\item{Z.W. Gortel, M.A. Za\l{}uska-Kotur, \L{}.A.~Turski, \refdo{{\em Phys. Rev. B} {\bf 52}, 16916 (1995)}{10.1103/PhysRevB.52.16920}}

\item{S. Dattagupta, \L{}.A. Turski, \refdo{{\em Phys. Rev. A} {\bf 32}, 1439 (1985)}{10.1103/PhysRevA.32.1439}}

\item{S.Q. Nguyen, L.A. Turski, \refdo{{\em Physica A} {\bf 290}, 431 (2001)}{10.1016/S0378-4371(00)00449-0}} 

\item{S.Q. Nguyen, L.A. Turski, \refdo{{\em J. Phys. A} {\bf 34}, 9281 (2001)}{10.1016/j.physa.2008.09.026}}

\item{M. Materassi, E. Tassi, \refdo{{\em Physica D} {\bf 241}, 729 (2012)}{https://doi.org/10.1016/j.physd.2011.12.013}}

\item{A. Kaufman, \refdo{{\em Phys. Lett. A} {\bf 100} (1984)}{10.1016/0375-9601(84)90634-0}}

\item{A. Kaufman, \refdo{{\em Phys. Lett. A} {\bf 109}, 87, (1985)}{10.1016/0375-9601(85)90261-0}}

\item{P.J. Morrison, \refdo{{\em Phys. Lett. A} {\bf 100}, 423 (1984)}{10.1063/1.4982054}}

\item{A.N. Kaufman, L.A. Turski, \refdo{{\em Phys. Lett. A} {\bf 120}, 331 (1987)}{10.1016/0375-9601(87)90725-0}}

\item{C.P. Enz, L.A. Turski, \refdo{{\em Physica A} {\bf 96}, 369 (1979)}{10.1016/0378-4371(79)90002-5}}

\item{Ching Lok Chong, \refdo{{\em J.Non-Newtonian Fluid Mech.} {\bf 292}, 104537 (2021)}{10.1016/j.jnnfm.2021.104537}}\nopagebreak

\item{J.A. Holyst, L.A. Turski, \refdo{{\em Phys. Rev. A} {\bf 45}, 6180 (1992)}{10.1103/PhysRevA.45.6180}}

\item{T.B. Boykin, \refdo{{\em Am. J. Phys.} {\bf 69}, 793 (2001)}{10.1119/1.1344169}}

\item{S.P.C. Martin, E.D. Siggia, H.A. Rose, \refdo{{\em Phys. Rev. A} {\bf 8}, 423 (1973)}{10.1103/PhysRevA.8.423}}
\end{enumerate}
\end{document}